\documentclass{IEEEtran}
\usepackage{textcomp}
\usepackage{ulem}
\usepackage{booktabs}
\usepackage{balance}
\usepackage{cite}
\usepackage[pdftex]{graphicx}
\usepackage{color}
\usepackage{import}
\usepackage{booktabs}
\usepackage{multirow}
\usepackage{amsmath}
\usepackage{ulem}
\usepackage{float}

\begin{document}

\title{Temporal Memory with Magnetic Racetracks}

% author names and affiliations
% transmag papers use the long conference author name format.

\author{\IEEEauthorblockN{Hamed Vakili\IEEEauthorrefmark{1},
Mohammad Nazmus Sakib\IEEEauthorrefmark{2},
Samiran Ganguly\IEEEauthorrefmark{2}, 
Mircea Stan\IEEEauthorrefmark{2}, 
\\Matthew W.~Daniels\IEEEauthorrefmark{3},
Advait Madhavan\IEEEauthorrefmark{3}\IEEEauthorrefmark{4},
Mark D. Stiles\IEEEauthorrefmark{3} and
Avik W. Ghosh\IEEEauthorrefmark{1,2}},\\
\IEEEauthorblockA{\IEEEauthorrefmark{1}Department of Physics,
University of Virginia, Charlottesville, VA 22903 USA}\\
\IEEEauthorblockA{\IEEEauthorrefmark{2}Department of Electrical and Computer Engineering, University of Virginia, Charlottesville, VA 22903 USA}\\
\IEEEauthorblockA{\IEEEauthorrefmark{3}Alternative Computing Group, National Institute of Standards and Technology, Gaithersburg, MD}\\
\IEEEauthorblockA{\IEEEauthorrefmark{4}Institute for Research in Physics and Applied Electronics, University of Maryland, College Park, MD}}
% <-this % stops an unwanted space

% The paper headers
%\markboth{Journal of \LaTeX\ Class Files,~Vol.~14, No.~8, August~2015}%
%{Shell \MakeLowercase{\textit{et al.}}: Bare Demo of IEEEtran.cls for IEEE Transactions on Magnetics Journals}
% The only time the second header will appear is for the odd numbered pages
% after the title page when using the twoside option.
% 
% *** Note that you probably will NOT want to include the author's ***
% *** name in the headers of peer review papers.                   ***
% You can use \ifCLASSOPTIONpeerreview for conditional compilation here if
% you desire.

% If you want to put a publisher's ID mark on the page you can do it like
% this:
%\IEEEpubid{0000--0000/00\$00.00~\copyright~2015 IEEE}
% Remember, if you use this you must call \IEEEpubidadjcol in the second
% column for its text to clear the IEEEpubid mark.

% use for special paper notices
%\IEEEspecialpapernotice{(Invited Paper)}

% for Transactions on Magnetics papers, we must declare the abstract and
% index terms PRIOR to the title within the \IEEEtitleabstractindextext
% IEEEtran command as these need to go into the title area created by
% \maketitle.
% As a general rule, do not put math, special symbols or citations
% in the abstract or keywords.

\maketitle
%\IEEEtitleabstractindextext{%
\begin{abstract}
Race logic is a relative timing code that represents information in a wavefront of digital edges on a set of wires in order to accelerate dynamic programming and machine learning algorithms. Skyrmions, bubbles, and domain walls are mobile magnetic configurations (solitons) with applications for Boolean data storage. We propose to use current-induced displacement of these solitons on magnetic racetracks as a native temporal memory for race logic computing. Locally synchronized racetracks can spatially store relative timings of digital edges and provide non-destructive read-out. The linear kinematics of skyrmion motion, the tunability and low-voltage asynchronous operation of the proposed device, and the elimination of any need for constant skyrmion nucleation make these magnetic racetracks a natural memory for low-power, high-throughput race logic applications.
\end{abstract}

% Note that keywords are not normally used for peerreview papers.
% make the title area
\maketitle

% To allow for easy dual compilation without having to reenter the
% abstract/keywords data, the \IEEEtitleabstractindextext text will
% not be used in maketitle, but will appear (i.e., to be "transported")
% here as \IEEEdisplaynontitleabstractindextext when the compsoc 
% or transmag modes are not selected <OR> if conference mode is selected 
% - because all conference papers position the abstract like regular
% papers do.
\IEEEdisplaynontitleabstractindextext
% \IEEEdisplaynontitleabstractindextext has no effect when using
% compsoc or transmag under a non-conference mode.

% For peer review papers, you can put extra information on the cover
% page as needed:
% \ifCLASSOPTIONpeerreview
% \begin{center} \bfseries EDICS Category: 3-BBND \end{center}
% \fi
%
% For peerreview papers, this IEEEtran command inserts a page break and
% creates the second title. It will be ignored for other modes.
\IEEEpeerreviewmaketitle

\section{Displacement based Magnetic Memories and Arrival Time Codes}
\label{sec:intro}

When energy efficiency becomes the predominant metric in computing systems, the choice of information representation also becomes important. A recently proposed temporal coding scheme, known as race logic \cite{madhavan_race_2014,RaceTrees, madhavan20174}, can have orders of magnitude energy improvements over classical approaches. In race logic, information is encoded in the relative timing between digital rising edges on different wires. This allows conventional Boolean primitives to perform non-traditional operations at a very low energy cost. Computations are generally performed by setting up the problem in a spatially arranged network of operators, like AND gates, OR gates, and temporal delay elements. Digital temporal wavefronts are presented to the inputs of such an array and the way the wavefront navigates through the network performs the computation. For example, an OR gate determines the first arriving signal, essentially performing a MIN function on the signals encoded on two wires. This approach is very efficient for dynamic programming problems like decision trees \cite{RaceTrees} or genetic sequencing \cite{madhavan20174}.

One major impediment in implementing race-logic-based temporal computing systems is the need for a memory that can easily store such temporally coded information.  Such storage would enable more complicated processing than can be done with simple logic gates. Though digital wavefronts can be generated and recorded with conventional Boolean circuits, conversions between temporal and binary representation incur a sizeable area and energy cost and hence limit the kinds of computations that can be performed. 

Magnetic devices play key roles in data storage from magnetic tapes to hard disk drives to tunnel junction memories. A racetrack memory \cite{parkin_magnetic_2008} is similar to a magnetic disk but without physically moving parts. Translating the magnetic configuration along the track plays the role of moving a magnetic tape. Racetracks with skyrmions~\cite{tomasello_strategy_2015}, a particular magnetic configuration discussed in Sec.~\ref{sec:dw-skyrmion} can be used to store Boolean information with the presence of a skyrmion at a particular location indicating a one and the absence a zero, for example. This information can be subsequently read by translating the skyrmions past a detector that reads the changing magnetic state through a resistance change. Translation can be achieved by several mechanisms including passing a current through a heavy metal layer underlying the track.  This current injects a spin current into the magnetic layer creating a spin-orbit torque \cite{hanneken_electrical_2015,camsari_stochastic_2017} that rotates the magnetization.  The subsequent local rotations of the magnetization give rise to an effective translation of the magnetization pattern.

\begin{figure}
    \centering
    \def\svgwidth{\linewidth}
    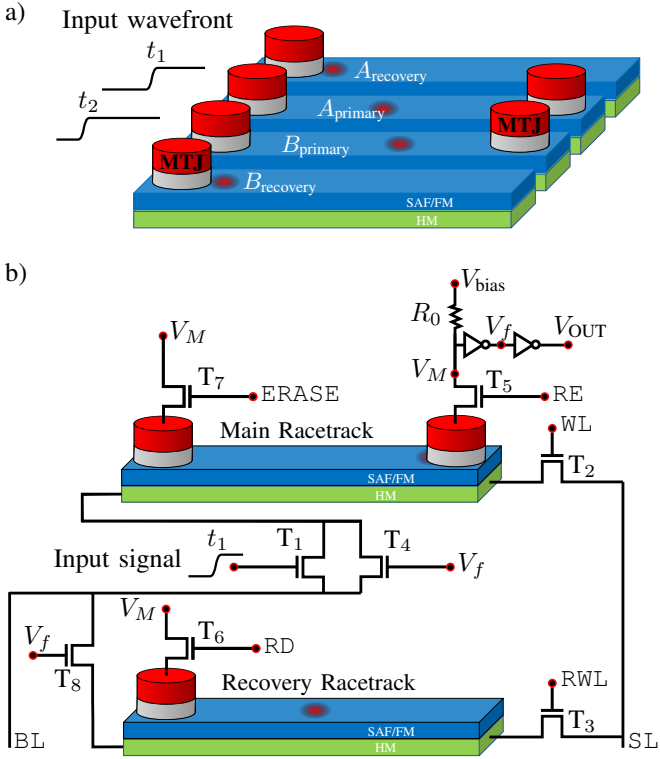
	
   \caption{Concept and circuit illustration: panel (a) shows the general arrangement of a single column of $N=2$ cells which can be used to store and play temporal wavefronts.  The racetracks consist of a bilayer of synthetic antiferromagnet (SAF) or a ferromagnet (FM) with magnetic tunnel junctions (MTJs) to detect when a skyrmion appears under them. Panel (b) shows a detailed description of the circuits in each cell.  Control lines consist of the bit line (BL), source line (SL),  word line (WL), read enable (RE), recovery line (RD),  recovery  word line (RWL), and erase line (ERASE).}
      \label{fig:CKT}
\end{figure}

In this paper, we present a design (Fig.~\ref{fig:CKT}) for a memory cell that converts information from the time domain to a displacement domain by using current pulses of varying lengths in time to translate skyrmions variable distances in space, thereby encoding the arrival times of the pulses. The displacement of the skyrmions is induced by applying a current along the racetrack, causing the skyrmions to move along the racetrack with a fixed velocity. During the write operation, the different timings of arriving digital edges on the wires lead to different lengths of current pulses and hence different translations of the skyrmions in the corresponding race tracks. When this memory needs to be read out, the temporal reference signal is provided to the array initiating current flow along the racetrack displacing the stored skyrmion. When the skyrmion reaches the end of the racetrack, its arrival triggers an output edge. We use pairs of racetracks to enable a non-destructive read as discussed in Sec.~\ref{sec:rl-memory}. 

The linear dependence of displacement on temporal difference, coupled with the possibility of variable operating speed, make a skyrmion-based memory very attractive for temporally coded systems.
A previously proposed temporal memory~\cite{madhavan_iscas2020} is based on memristors, which have logarithmic responses to inputs. These are complicated to use in exact timing codes because the nonlinear response has to be actively managed, and restricts operation to only a small dynamic range of the device.  The development of an energy and space efficient way to store temporal signals with linear read/write dynamics greatly expands the range of algorithms that can be addressed with race logic.

We present details of this memory, the non-destructive readout, a way to reset memory, and a comparison with conventional Boolean approaches in the next few sections. Section~\ref{sec:dw-skyrmion} provides background on the magnetic technology that is used. Section~\ref{sec:rl-memory} describes how complementary metal-oxide-semiconductor (CMOS) circuits can be interfaced with such technology to perform read and write operations. Section~\ref{sec:resdis} describes our simulation results for ideal skyrmion racetracks followed by a discussion of non-idealities that arise as a result of imperfections in material properties and how they affect the the operation of this memory. 

\section{Domain Walls and Skyrmions for Memory}
\label{sec:dw-skyrmion}

Magnetic memories generally store information in the orientations of the microscopic magnetic domains that reside within the magnetic material. Magnetizations tend to have their energy minimized when the moments point in either direction along a preferred axis. Such a binary configuration naturally lends itself to binary information encoding in which one direction corresponds to 0 and the other to 1.  This encoding is used in magnetic tapes, hard disk drives, and magnetic random access memory. In continuous media, the magnetization tends to form domains with the magnetization in two neighboring regions roughly uniform and separated by a narrow region where the magnetization rotates from one domain to the other.  The region with the rotating magnetization is referred to as a domain wall.

When magnetic materials are fabricated in a 2D geometry, energetic considerations typically prefer that the magnetization lie in the plane. However, it is possible to tune the anisotropies of the material such that the magnetization tends to point parallel to the interface normal, which we label $\hat z$.  We choose such a material in this work. In the uniform configuration, the magnetization points in the same direction everywhere in a sample, either along $\hat z$ or $-\hat z$. However, other configurations can exist in a metastable state~\cite{belavin1975metastable}. One such configuration consists of two stable regions (one along $\hat z$ and one along $-\hat z$) separated by a localized 180$^\circ$ domain wall.  Localized domain walls are generated between oppositely oriented magnetic regions through the balance of exchange and anisotropy forces, the latter of which can arise due to intrinsic magnetocrystalline anisotropy or from magnetostatic interactions~\cite{thiaville_domain_2002,beach_dynamics_2005}.

The domain wall we have just described divides two regions of uniform magnetization. A skyrmion can be imagined by making the domain wall circular to surround a  region of uniform magnetization (inside the circle) from an region (outside the circle).  For skyrmions, the domain wall separating the two regions twists in a particular way, which is described by a topological index known as a winding number. Using ultra-thin ferromagnets (FM) on top of heavy metal (HM) layers breaks inversion symmetry and generates an additional energy term, the Dzyaloshinskii-Moriya interaction (DMI) \cite{dzyaloshinsky_thermodynamic_1958}, which stabilizes skyrmions.

Both domains and skyrmions have been proposed as fundamental elements of future magnetic memories.  For domains, ones and zeros are encoded in different magnetization directions; for skyrmions, it is their presence or absence which encodes a bit. In the particular class of memory devices called racetracks~\cite{tomasello_strategy_2015,parkin_magnetic_2008} that we consider here, the sample geometry is a long thin wire. The information stored in the racetrack can be translated along the wire by passing a current through the wire. Here, this translation is facilitated by the inclusion of a heavy metal layer next to the magnetic layer. An electric current driven through the heavy metal layer injects a spin current, through a spin Hall effect, perpendicular to the FM/HM interface into the magnetic layer, thereby applying a spin-orbit torque to the ferromagnetic layer~\cite{sampaio2013nucleation,tomasello_performance_2017,zhang_magnetic_2016,vakili_self_2020}. This torque can drive skyrmions and domain walls in a variety of ultra-thin  magnetic systems at speeds as high as 500~m/s to 1~km/s for relatively low current densities. Possible systems include ferromagnets, antiferromagnets (AFMs), heavy-metal heterostructures, and the two systems of particular interest here: synthetic antiferromagnets~\cite{legrand_room-temperature_2019,dohi_formation_2019,xie_computational_2019}, and nearly compensated ferrimagnets. In our simulations, we consider a racetrack that is either a two-sublattice near-compensated ferrimagnet (FiM) like CoGd or a synthetic antiferromagnet, where the low saturation magnetization gives small-sized high-speed skyrmions that propagate along the track without deflection to the side as discussed in Sec.~\ref{sec:resdis}.

The dynamics of domain walls and skyrmions under the effect of an electronic current is well studied~\cite{thiele_steady-state_1973,tveten_2013}, and can be modeled using classical mechanics. In this work we make use of these classical kinemetic behaviors of domain walls and skyrmions by exploiting the simple linear response $x=vt$ between displacement $x$ and time $t$, mediated by a constant velocity $v$. By using a constant, known-amplitude current pulse to drive a skyrmion at $v$ for some period of time $t$, we find that $t$ is stored in the position $x$ of the skyrmion along the track, and $t$ can be reliably recovered simply by driving the skyrmion in reverse and waiting for however long it takes the skyrmion to return to its origin point on the track.

Both domain walls and skyrmions have topological protection of different types, which means that there is a significant barrier to eliminating them. 
This topological protection gives them some of the stability needed for long lifetimes 
%and rigidity to be unpinned from possible defects in a racetrack, especially at lower speeds where they can deform their shapes 
\cite{bogdanov_chiral_2001}. 
% ^ Is this true? Aren't DWs exceptionally susceptible to pinning?
All of these properties make  domain walls and skyrmions good candidates for racetrack memories. We choose skyrmions for simulations, but most of the work here applies equally well  to domain walls and other possible magnetic solitons. We choose synthetic antiferromagnetic systems because their motion has useful features as explained below.

To model skyrmion motion in synthetic antiferromagnets, we use a collective coordinate description of a rigid skyrmion, that is, we assume the skyrmion texture has only translational degrees of freedom and derive equations for these from the governing Landau-Lifshitz-Gilbert equations. In antiferromagnets and synthetic antiferromagnets, the resulting rigid soliton dynamics is given by a second order differential equation~\cite{tveten_2013}, characterized by an effective skyrmion mass. The effective mass of a rigid skyrmion in a synthetic antiferromagnet is inversely proportional to the interlayer exchange coupling~\cite{daniels_2019}. We assume very strong coupling, which sends the mass to zero and reduces the dynamics to a first-order description. We also assume that the only driving forces on the skyrmions come from spin orbit torques arising from the spin Hall effect (spin Hall angle $\Theta_\text{sh}$) in the heavy metal layer. 

The resulting Thiele equation gives an instantaneous speed for the skyrmion  \cite{buttner_theory_2018,thiele_steady-state_1973}
\begin{equation}
    v =\frac {\pi\gamma\hbar}{2e}\frac{I_d\Delta}{\sqrt{(4\pi) ^2 \langle N\rangle^2+\alpha^2{\mathcal{D}_{xx}^2} }}\frac{\Theta_{\mathrm{sh}}}{\Sigma_i t_i M_{s_{i}}}j,
\label{eq:mobility}
\end{equation}
where $j$ is the electrical current density in the heavy metal layer, $t$ is the thickness of that layer, $\Delta$ is the characteristic domain wall length, $M_s$ is the saturation magnetization in each layer of the synthetic antiferromagnet, $\alpha$ is the Gilbert damping, and $\gamma$ is the gyromagnetic ratio. Two characteristic properties of the skyrmion texture are the (integer) winding number
\begin{equation}
\langle N\rangle = \frac{1}{4\pi}\int \text{d}x\,\text{d}y\; \mathbf{m} \cdot 
\left(\frac{\partial \mathbf{m}}{\partial x} \times
\frac{\partial \mathbf{m}}{\partial y}\right)
\end{equation}
and the longitudinal component of the dissipation tensor 
\begin{equation}
\mathcal{D}_{xx} = \int \text{d}x\,\text{d}y\,(\partial_x \boldsymbol{m})^2,
\end{equation}
which provides a kinetic friction force for the moving skyrmion.  The direction of the skyrmion motion is determined by the ratio of $4\pi \langle N\rangle$ to $ \alpha\mathcal{D}_{xx}$.  Because of the off-diagonal nature of a skyrmion's velocity response, $\alpha\mathcal{D}_{xx}$ ends up being principally responsible for the longitudinal skyrmion motion along the track while the winding number $\langle N\rangle$ gives a transverse Magnus force. Finally, the factor $I_d$ accounts for the spatially varying response of the skyrmion due to its spin texture; this term can be approximated as 
 \begin{equation}
     I_d \approx e^{-r/\Delta} + \frac{\pi r}{\Delta}
 \end{equation}
in rigid skyrmions of radius $r$, where $r$ is defined by the $m_z=0$ contour~\cite{buttner_theory_2018}. 

\begin{figure*}
   \def\svgwidth{\linewidth}
    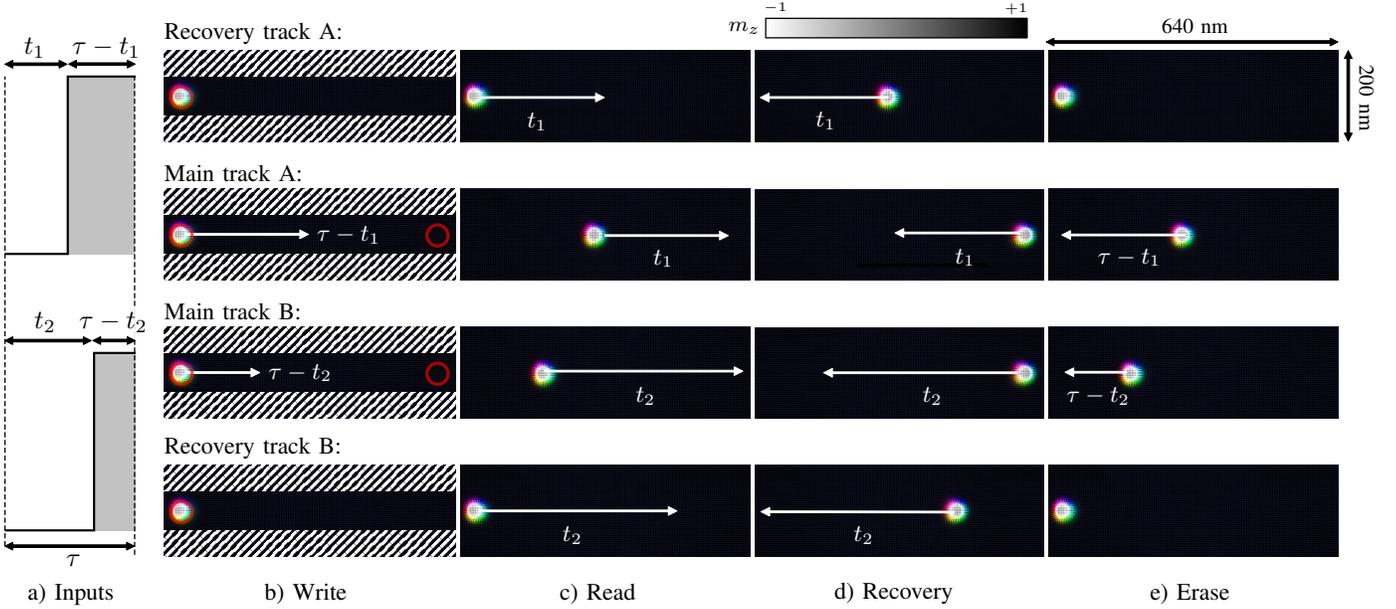
    \caption{(a) Two example waveforms presented as input to be stored by the memory cell. (b-e) Four operational phases of the racetrack temporal memory from a micromagentic simulation. The figures show a background of up-spin magnetic material (black) with a down-spin skyrmion (white); color indicates the in-plane spin orientation. In each phase, the skyrmion location is shown at the beginning of that phase while the arrows point to the location of the skyrmion at the end of the phase. The red circles in the first column indicate the magnetic tunnel junction (MTJ) placement, and the patterned area of the racetracks represents an energy barrier.  Amplified repulsion from edges helps keep the skyrmions on their path to the read MTJ locations while preventing them from annihilation or pinning at the edges.}
\label{fig:skyrmion-snapshots}
\end{figure*}

Note that, contrary to ferromagnetic skyrmions, skyrmions in synthetic antiferromagnets are not expected to experience a Magnus force, due to cancellation between the two oppositely magnetized layers~\cite{barker2016}. This cancellation implies that $\langle N\rangle=0$, reducing Eq.~\eqref{eq:mobility} to
\begin{equation}
    v=\frac {\pi\gamma\hbar}{2e}\frac{ I_d\Delta }{\alpha{\mathcal{D}_{xx}} }\frac{\Theta_{\mathrm{sh}}}{\Sigma_i t_iM_{s_{i}}}j .
\label{eq:simmobility}
\end{equation}
We develop a circuit module that integrates Eq.~\eqref{eq:simmobility} over time to obtain the instantaneous skyrmion location on the racetracks, which is used to capture the effect of micromagnetic simulations as shown in Fig.~\ref{fig:skyrmion-snapshots}.
%\begin{figure*}
%   \centering
  % \def\svgwidth{\linewidth}
    %\import{figs/}{Fig2.pdf_tex}
    %\caption{Four operational phases of the racetrack temporal memory from a micromagentic simulation. The figures show a background of up-spin magnetic material (black) with a down-spin skyrmion (white); color indicates the in-plane spin orientation. In each phase, the skyrmion location is shown at the beginning of that phase while the arrows point to the location of the skyrmion at the end of the phase. The red circles in the first column indicate the MTJ placement.}%
%    \label{fig:micromagnetics}
%    \label{fig:skyrmion-snapshots}
%\end{figure*}

In memories based on a classical Boolean representation, a skyrmion is used to implement a binary \texttt{1} or a \texttt{0} by virtue of its presence or absence. This procedure requires nucleating and destroying skyrmions~\cite{sampaio2013nucleation}, which costs substantial energy. It can also be difficult to engineer in a deterministic fashion. The energy cost is necessary, because the resulting energy barrier that must be overcome provides the necessary stability for a skyrmion to be used as a memory. In our memory, we just translate skyrmions without creating or destroying, aside from initializing them for the first time in the device. Once the skyrmions are nucleated, it is their position on the racetrack that encodes information and hence a ``0'' state can be stored by returning the skyrmion to its initial position, instead of destroying it. This is made possible by the long lifetimes and relatively low currents needed %\todo{(write approx magnitudes is brackets)}
to drive skyrmions. 

The detection scheme is based on the tunneling magnetoresistance (TMR) found in magnetic tunnel junctions.  In a magnetic tunnel junction, two magnetic layers are brought close to each other and separated by an insulating layer. The resistance through this material stack depends on the relative orientations of the magnetizations on either side of the insulator. This effect allows the information stored in these magnetic domains to be electrically read out. In this system, the insulating layer is deposited directly on top of the racetrack layer and an additional ferromagnetic layer with a fixed magnetization is deposited on top of that. The resistance measured by this detector changes when there is a skyrmion in the race track below the rest of the tunnel junction.

% The reading of stored bits on the racetrack is a two step process (see fig. \ref{general}d,e). At first, a skyrmion on the `main racetrack' is driven by a uniform current till the end of the race track. We couple this main track with a companion `recovery racetrack' and during the read process, we move the recovery skyrmion from its initial position along with the main skyrmion. Then after reading the main skyrmion using a magnetic tunnel junction (MTJ) located at the end of the racetrack, the skyrmions are moved back to their original positions, with the help of the companion located in the recovery racetrack by reversing the drive current. The duration of this restoration pulse is provided by reading the recovery skyrmion using a MTJ placed at the beginning of the racetrack. When the skyrmions reach the end of a racetrack, the information about their relative position on that racetrack is lost. Therefore, the recovery racetrack which preserves the relative positions of main skyrmion by storing its position temporarily preventing the loss of information after each reading process.

%\begin{figure}[t!]
%\centering
%\includegraphics[scale=0.5]{ckt.JPG}
%\caption{Proposed race logic architecture of each processing unit, a racetrack and corresponding recovery track.}
%\label{circuit}
%\end{figure}

\section{Temporal Memory Circuits and Simulations}
\label{sec:rl-memory}
Two cells of the proposed magnetic skyrmion based architecture are shown in Fig. \ref{fig:CKT}a. The full architecture consists of $N$ cells, each comprising two parallel racetracks. The memory is interfaced with an $N$ channel data bus on which the input wavefronts arrive. Our temporal memory allows for these $N$-channel wavefronts to be stored in these cells, in parallel as a single write operation. These written values can also be replayed back in a single parallel read operation. As we discuss below, this reading process destroys the information in a single racetrack. The second racetrack for each of the $N$ channels retains the stored information allowing a recovery operation to restore the previously read state. Finally, an erase operation returns the cells to a zero value.  All operations involve translating a single preexisting skyrmion in each of the racetracks so that none are ever created or destroyed. Note that, between these operations, all the drive signals can be turned off without losing the data, due to the inherent non-volatility of the skyrmions.  We present these four operations next, combined with the details of the temporal to spatial mapping scheme. The bias voltages for  these operations are given in Table~\ref{tab:bias-conditions}.

\subsection{Temporal-to-Spatial Mapping Scheme}

The linearity of skyrmion displacement with the drive current pulse length is central to storing temporal data into a linearly mapped spatial skyrmion arrangement on a racetrack. Defects in the racetrack can give rise to non-linear effects, which we discuss in Sec.~\ref{sec:resdis}. Assuming a defect-free fabrication process, the racetracks have a finite capacity to store the temporal information, which depends on the storable racetrack length $L$, given by the total time, $\tau=L/v$, needed by the skyrmion to travel with velocity $v$ from the starting point to the end point of the racetrack. This temporal memory architecture requires that during the `read' process, the arrival time of a skyrmion to the read magnetic tunnel junction (MTJ) located at the end of the racetrack corresponds to the stored temporal data $t_1$.

To meet this requirement, the computation window for the cell is set as the transit time $\tau = L/v$. The time $t_1$ is the arrival time of a rising edge, the time delay after which the wavefront voltage switches from low to high. Provided that $0 \le t_1 \le \tau$, the write process to store that value requires displacing the skyrmion for $\tau-t_1$ after the rising edge arrives. This displacement positions the skyrmion properly to represent the wavefront arrival. The fidelity of this operation is ensured because the CMOS-based control circuitry turns off the drive current sufficiently quickly to minimize any overshoot of skyrmion on the racetrack. 

We incorporate a read-enable signal whose charging up corresponds to the temporal axis origin for the read process, $t = 0$. This read-enable signal restarts the drive current as well as the read MTJs reporting the temporal data $\tau-t_1$ by toggling the read voltage caused by skyrmion/wavefront arrival. The time of arrival for the skyrmion to the read sensing MTJ is $t_1$, after which the pulse jumps up to high upon a read and stays high for the remainder of the clock cycle $\tau-t_1$, thus reproducing the incoming wavefront.  As usual, the fidelity of the read process  is ensured by the much faster time scales of CMOS control circuitry compared to the time scale of the current pulses used to move the skyrmions.

\begin{figure}%
   \def\svgwidth{\columnwidth}
    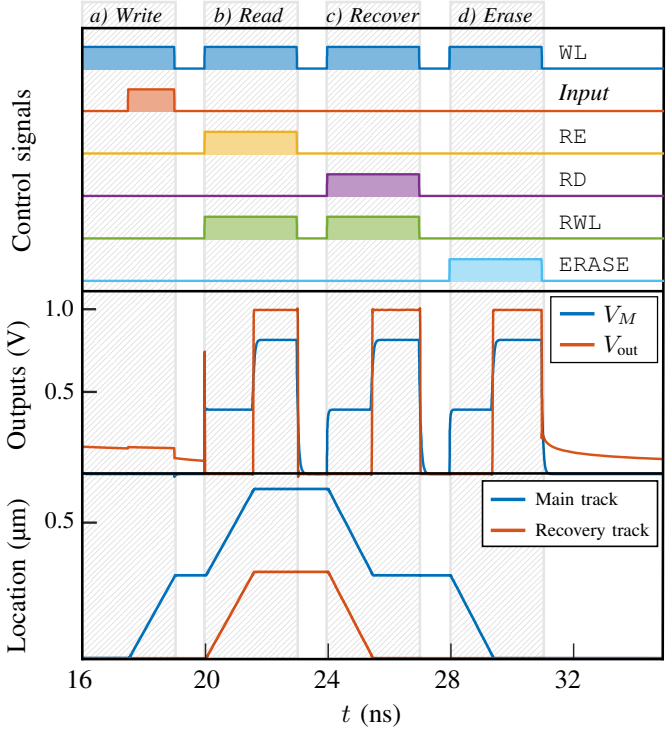
    %\import{four-phase-operation.pdf_tex}
%\includegraphics[width=\textwidth]{Ckt_operation}%
\caption{Control signals, output signals, and skyrmion positions on the two tracks through four phases memory cell operation.  Top: control bits for the write line (\texttt{WL}), read enable (\texttt{RE}), recovery line (\texttt{RD}),  recovery write line (\texttt{RWL}), erase line (\texttt{ERASE}), and the incoming signal. The middle panel shows the synchronizing signal $V_\text{OUT}$ and the node voltage $V_M$. The bottom panel shows the instantaneous skyrmion positions on the tracks. The four operational phases are highlighted with gray boxes. a) \texttt{WL} drives the skyrmion along the main racetrack, once the input arrives. b) \texttt{WL} and \texttt{RWL} drive the main and recovery skyrmions, respectively, until the main track skyrmion reaches the MTJ read stack, which is enabled by RE. The resulting resistance change in the MTJ, detected via the feedback voltage $V_f$ (inverse of $V_{out}$), halts the drive currents through the tracks. c) \texttt{WL} and \texttt{RWL} drive the main and recovery skyrmions as in the read operation, but with the relative polarity of  source  and bit lines inverted to enable a reverse drive current. Motion stops when the recovery track skyrmion is detected by the \texttt{RD}-enabled MTJ stack, restoring the original position of the main track skyrmion before the read operation. d) The \texttt{ERASE} signal drives the main skyrmion back to the origin point until it is detected and the drive current is turned off.
}%
\label{fig:cktop}%
\end{figure}

\subsection{Write Operation}
The circuit for the write operation is shown in fig. \ref{fig:CKT}(b). As described before, we assume pre-nucleated skyrmions in the starting position of the main racetrack and the recovery track as the initial condition for the write operation. Note that the write operation involves only the primary/main track and not the recovery track, hence only transistors $\text{T}_{1}$ and $\text{T}_{2}$ are turned on. All the other transistors are in the off state. The write operation corresponds to shifting the skyrmion in the primary racetrack, from its starting position to its final position, by the application of the write current {(fig. \ref{fig:skyrmion-snapshots}c.)}. The final position of the skyrmion depends on the duty cycle of the incoming wavefront. This is performed by applying high ($V_\text{DD}=1$~V) voltage to the bit line (BL) and ground to the source line (SL) to generate the current, while the write line (WL) transistor, $\text{T}_{2}$, is turned on to establish a current path. 

When the incoming temporally coded high signal arrives at the gate of the transistor $\text{T}_{1}$, the write path through the heavy metal layer is turned on and the skyrmion begins its motion along the track. When the incoming signal falls to ground, the transistor $\text{T}_1$ turns off, causing the skyrmion to stop moving and isolating the main racetrack from the BL. This isolation of the main racetrack from the BL ensures no skyrmion movement after the end of write operation. This captures the incoming high temporal signal, and converts it into a spatial displacement of the skyrmion in the track. The control signals and corresponding skyrmion motion as a function of time are shown in Fig.~\ref{fig:cktop}a.

\subsection{Read Operation}

Once the input signal is captured in the racetrack by displacement of the skyrmion to the appropriate position, it can be read. This read operation is a destructive operation because once the track is read out, the position information of the skyrmion on that track is lost. To overcome this issue, we use a corresponding recovery track, which operates as shown in Fig.~\ref{fig:CKT}b. The time period of the applied bias voltage has to be same in both tracks to ensure data integrity. The BL is biased to high ($V_\text{DD}=1$~V) and SL is biased to ground, similar to the write operation.  Transistor $\text{T}_1$ is turned off and is not used in the read operation. The node voltage $V_{M}$ is low at the beginning of the read operation due to the low resistance value of the readout MTJ in the absence of skyrmion. 

The amplified node voltage  $V_f$ is high and turns on the transistors $\text{T}_4$ and $\text{T}_8$.    This sets the system up so that the same current flows through both the primary and recovery track, as soon as the rising edge that initiates the read operation appears at the inputs to transistors $\text{T}_2$ and $\text{T}_3$. Moreover, the aspect ratio of the transistors $\text{T}_1$, $\text{T}_4$, and  $\text{T}_8$ are adjusted to make sure that the same current flows through the main racetrack during both the write and read operation and  the recovery racetrack during the read operation. The same current through the main and recovery racetracks ensures the same duty cycle of the stored and retrieved wavefront.  

When the rising edge arrives, the skyrmions on both the primary and recovery tracks begin to move in the same direction until the skyrmion in the primary track reaches the MTJ at the end of the track (see fig. \ref{fig:skyrmion-snapshots}c. for a micromagnetic simulation and  fig. \ref{fig:cktop}b. for a circuit simulation). This results in an output edge, corresponding to time $\tau-t_1$, which is the correct stored value. The important point is that since the main and recovery tracks see the same current pulse, they exhibit identical dynamics, hence keeping the information that would have otherwise been lost. To ensure that this information is stored correctly to perform successive read operations, the recovery track has to be turned off as soon as the output triggers a rising edge. This is achieved by nMOS transistors $\text{T}_8$ and $\text{T}_4$ which turn both current paths off as the $V_f$ goes low when $V_{M}$ and $V_{OUT}$ trigger a rising edge. Note that the CMOS time constants are much faster than the magnetic dynamics of skyrmion movement and hence can be treated as instantaneous.  

In order to correctly detect the skyrmion, we use MTJ readout circuitry based on the difference in resistance between there being a skyrmion beneath the detector, $R_\text{sk}$, or not, $R_\text{P}$.  This difference can be characterized by an effective TMR, $\beta=(R_\text{sk}-R_\text{P})/R_\text{P}$.  A bias voltage $V_\text{bias}$ with a reference resistor $R_{0}$ produces a voltage swing that depends on the effective TMR $\beta$ at node $V_{M}$ in the presence or absence of a skyrmion. The voltage at the node $V_{M}$ is then amplified by an inverter to feed high or low feedback voltage $V_f$ to the gates of the nMOS transistors $\text{T}_{4}$ and $\text{T}_{8}$. Note that the readout circuitry is biased in such a pattern that the feedback voltage $V_f$ turns on the transistors $\text{T}_{4}$ and $\text{T}_{8}$ during the read, recovery, and erase operations only.   

The reference resistance should provide the maximum contrast between the skyrmion and no-skyrmion states under the MTJ. To ensure functionality, $R_0$ must be between $R_{0-}$ and $R_{0+}$, where
\begin{align}
R_{0\pm} &= \frac{R_\text{sk}}{2V_\text{min}}
\frac{\beta}{\beta+1}
\Bigg[V_\text{bias} - V_\text{min} (1+2/\beta)\nonumber\\
    &\; \pm \sqrt{V_\text{min}^2 -2V_\text{bias}V_\text{min}(1+2/\beta)+V_\text{bias} ^2}\;\Bigg].
\end{align}
Here, $V_\text{min}$ is the minimum voltage difference between the skyrmion and no-skyrmion states needed at node $V_M$ to ensure switching behavior at $V_f$, and $V_\text{bias}$ is the bias voltage. The maximum possible voltage swing is given when $R_0 = \sqrt{R_\text{sk}R_\text{P}}$.
Higher resistances increase efficiency and can be achieved by using a thicker MgO layer in the MTJ.  The TMR used in this expression, $\beta$, is reduced from the bare TMR by the fill factor of the skyrmion underneath the MTJ due to the non-uniform magnetization in the skyrmion. It is essential that this effective TMR be large and the filling of the MTJs by the skyrmions be almost complete because the read-out circuitry begins to function poorly when the effective TMR is below 50~\%. Using domain walls instead of skyrmions would give a larger read-out signal at the potential cost of more complicated motion.

Though $V_M$ is connected to three different MTJs, only one of them is active in each operating phase. This node and its associated readout circuitry can therefore be shared by these MTJs, reducing the area and energy cost associated with static read currents. Such selective operation of the readout circuitry is performed by transistors $\text{T}_5$, $\text{T}_6$, and $\text{T}_7$. In the read operation, for example, only $\text{T}_5$ is active, while $\text{T}_6$ and $\text{T}_7$ are turned off.

%$R_\text{MTJ}$ is the anti-parallel resistance, $\beta$ is the TMR, $\Delta V_M$ is the minimum operable voltage difference needed at node $V_M$, $V_bias$ is bias voltage. Higher $R_\text{MTJ}$ is preferable as it increases efficiency, this can be achieved by using a thicker MgO layer in the MTJ.  The TMR used in this expression, $\beta$, should be understood as the difference in resistance of the MTJ when there is a skyrmion below it and when there is not.  This value will be reduced from the bare tunneling magnetoresistance by the effective fill factor of the skyrmion underneath the MTJ.

%\begin{table*}
%\centering
%\caption {BIAS VOLTAGE CONDITION FOR RACE LOGIC AND MEMORY OPERATIONS}
%\begin{tabular}{ |c| c| c| c| c| c| c| c| c|}
%\hline
 %& BL & SL & WL & RWL & RE (read enable) & RD (read disable) & Reset \\ 
%\hline
%Write & $V_{SHIFT}$ & 0 & $V_{DD}$ & 0 & 0 & 0 & 0 \\ 
%\hline
%Read & $V_{SHIFT}$ & 0 & $V_{DD}$ & $V_{DD}$ & $V_{DD}$ & 0 & 0 \\ 
%\hline
%Recover & -$V_{SHIFT}$ & 0 & $V_{DD}$ & $V_{DD}$ & 0 & $V_{DD}$ & 0 \\  
%\hline
%Reset & -$V_{SHIFT}$ & 0 & $V_{DD}$ & 0 & 0 & 0 & $V_{DD}$ \\ 
%\hline
%\end{tabular}
%\end{table*}
\begin{table}

\centering
\caption {Bias voltages for race logic and memory operations. Columns correspond to the labeled circuit nodes in Fig.~\ref{fig:CKT}b.}
\begin{tabular}{@{}lrrrrrrr@{}}
\toprule
 & \texttt{BL} & \texttt{SL} & \texttt{WL} & \texttt{RWL} & \texttt{RE} & \texttt{RD} & \texttt{ERASE} \\ 
\midrule
Write & $V_\text{DD}$ & 0 & $V_\text{DD}$ & 0 & 0 & 0 & 0 \\ 
Read & $V_\text{DD}$ & 0 & $V_\text{DD}$ & $V_\text{DD}$ & $V_\text{DD}$ & 0 & 0 \\ 
Recovery & 0 & $V_\text{DD}$ & $V_\text{DD}$ & $V_\text{DD}$ & 0 & $V_\text{DD}$ & 0 \\  
Erase & 0 & $V_\text{DD}$ & $V_\text{DD}$ & 0 & 0 & 0 & $V_\text{DD}$ \\ 
\bottomrule
\end{tabular}
\label{tab:bias-conditions}
\end{table}

\subsection{Recovery Operation}

The third operation  which compensates for the destructive read, is the recovery operation. It restores the primary and recovery skyrmions to their original (i.e. before the read process) state. This operation is the opposite of the read operation and is performed almost identically, except with a  reversed applied current. Hence, the BL is biased to ground while the SL is biased to high ($V_\text{DD}$=1V), reversing the direction of current and hence the direction of motion of the skyrmions (fig. \ref{fig:skyrmion-snapshots}d.). Since the output MTJ in the recovery operation is the one at the beginning of the recovery track, transistors $\text{T}_5$ and $\text{T}_7$ can be disabled, while transistor $\text{T}_6$ is enabled. This allows the readout circuits to trigger when the skyrmion returns to its original location under the MTJ in the recovery track (fig. \ref{fig:cktop}c.). Note that as in the read operation, this MTJ detects the end of the operation and turns off transistors $\text{T}_8$ and $\text{T}_4$. After this operation is complete, the skyrmion in the recovery track is restored to its default origin position, while the skyrmion in the primary track is restored to its value before the read. The cell is now ready for another read as described before, or an erase operation. The recovery operation also replays the stored temporal data like the read operation (fig. \ref{fig:cktop}b) does.

\subsection{Erase Operation}

The final operation is the erase operation to return the cell to its configuration before any write operations, i.e., the configuration with the skyrmions on both racetracks at the origin.  The erase operation is similar to the write operation as only the skyrmion in the main track is moved, but the direction of the drive current and hence skyrmion motion is reversed (fig. \ref{fig:skyrmion-snapshots}e.). The ending of this phase is determined by the MTJ at the beginning of the main track. Note that in this phase, the other two MTJs are not required, causing transistors $\text{T}_5$ and $\text{T}_6$ to be disabled, while transistor $\text{T}_7$ can be enabled. As soon as the node voltage $V_f$ of the readout circuitry turns low, the transistor $\text{T}_4$ turns off, signaling the end of the erase operation (fig. \ref{fig:cktop}d.). The cells are now ready to write a new state. 

\section{Results and Discussion}
\label{sec:resdis}

In order to understand and quantify the performance of this wavefront memory cell, we perform detailed circuit simulations using a modular approach. In particular, we construct a complete circuit model using the $45$~nm technology node obtained from the Predictive Technology Model~\cite{ptm_model} for the driving transistors and the module discussed in Sec.~\ref{sec:dw-skyrmion} for the dynamics of the skyrmions in the magnetic racetracks.

To describe the magnetic dynamics, our simulations use a Gilbert damping constant $\alpha = 0.1$, saturation magnetization $M_s = 3\times 10^5 ~\text{A}/\text{m}$, magnetic anisotropy $K = 135~\text{kJ}/\text{m}^{3}$, Dzyaloshinskii-Moriya interaction strength $D = 1.2~\text{mJ}/\text{m}^{2}$, and exchange stiffness $A_{ex} = 7.5\times 10^{-12}~\text{J}/\text{m}$. Consequently, the domain wall width is $\Delta=\pi D/4K\approx 7.0$~nm and the skyrmion radius is approximately $24$~nm~\cite{Wang2018}. 
We use these parameters to model a synthetic antiferromagnet with thickness of $t_\text{FM} =2$~nm for each layer, length of $640$~nm, and width of $200$~nm. The heavy metal thickness is assumed to be $10$~nm. The MTJ diameter is $40$~nm with a TMR of $\approx 400$~\%, based on $R_P= 6.67$~k$\Omega$ and $R_{AP}= 33.3$~k$\Omega$.
The circuit behaves acceptably provided the bare TMR is greater than 100~\%, or an effective TMR of 50~\%.
With the chosen parameters, the skyrmion size is comparable to that of the MTJ.  From the structure of the skyrmion, we compute a 50~\% reduction in the TMR to an effective TMR close to 200~\%.  The reference resistance is taken to be $R_0=15$~k$\Omega$, giving a swing voltage of  $\approx$~425~mV. %We have checked t~

The saddle point barrier to skyrmion annihilation is $E_\text{b}\approx 23 A_\text{ex}t_\text{FM}-E_\text{skyr}\approx 100~ kT$, where $k$ is the Boltzmann constant and $T$ is the temperature. Assuming a mean lifetime to annihilation of \cite{bessarab_lifetime_2018} $\tau = f_0e^{E_\text{b}/k_BT}$ with $f_0 = 10^{-10}$, the lifetime of these skyrmions is in the range of years at room temperature. While the Magnus force is small in the compensated synthetic antiferromagnet, the wide track requires skyrmion injection right down the middle to hit the MTJ. This constraint can be avoided with narrower racetracks with repulsive edges.

%In our simulations, we take the Gilbert damping constant $\alpha = 0.1$, the saturation magnetization $M_s = 3\times 10^5 ~\text{A}\text{m}^{-1}$, the magnetic anisotropy $K = 190~\text{kJ}/\text{m}^{3}$, the Dzyaloshinskii-Moriya interaction strength $D = 1.2~\text{mJ}/\text{m}^{2}$, and the exchange stiffness $A_{ex} = 7.5\times 10^{-12}~\text{J}/\text{m}$. Consequently, $\Delta=\pi D/[4(K-\mu_0 M_s^2/2)]\approx 7.1$~nm and the skyrmion radius is approximately $42$~nm~\cite{Wang2018}. %$R = \pi D\big(A_{ex}/16 AK^2-\pi ^2D^2(K-\mu_0 M_s^2/2)\big)^{1/2}\approx 42$~nm .
%We use these parameters to model a synthetic antiferromagnet with thickness of $2$~nm for each layer, length of $640$~nm, and width of $200$~nm. The heavy metal thickness is assumed to be $10$~nm. The MTJ diameter is $40$~nm with a TMR of $\approx 200\,\%$. 
%\textcolor{red}{With the chosen parameters skyrmion size is comparable to the MTJ size so that the effective TMR for a skyrmion would not be small to make the detection too difficult. The lifetime of skyrmion for the chosen parameters would be in the years range at the room temperature based on estimation using the lifetime equation \cite{bessarab_lifetime_2018} $\tau = f_0e^{E_{barrier}/k_BT}$ with $f_0 = 10^{-10}$. [How is $E_\text{barrier}$ being computed to arrive at "years"? -MWD]}

\begin{figure}
      \def\svgwidth{\columnwidth}
    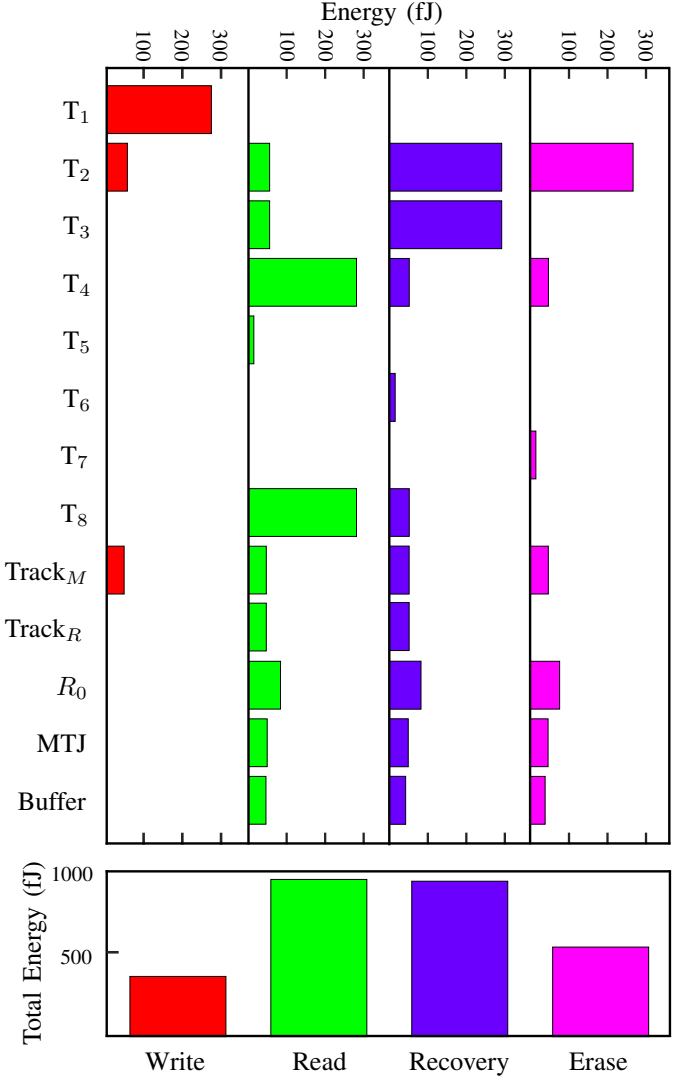
    % \import{bar-plots.pdf_tex}
    \caption{Total energy, and energy breakdown per circuit element, for an average case of wavefront capture on a set of 640~nm racetracks. The read and recovery processes consume the most energy because both tracks are active in these operations. In all operations, the largest energy dissipation arises from transistor leakage, with only minimal contributions from Joule heating in the main (Track$_M$) and recovery (Track$_R$) racetracks. }
    \label{fig:full_energy}
\end{figure}

Fig.~\ref{fig:full_energy} gives the energies consumed for these four operations in each component, for an average case in which the skyrmion is moved to the middle of the racetrack. The encoding of the wavefront arrival time is linear on the skyrmion motion. Transient simulations on our circuit allow us to capture the in-depth energy consumption in each component of the cell.   

The bottom panel of Fig.~\ref{fig:full_energy} shows that the energy consumption is highest in the read and recovery operations, as it involves driving both racetracks, while the write and erase operations, single track operations, consume relatively less energy. For the chosen parameters for the racetrack and the racetrack drive currents ($\approx 244$~{\textmu}A), the total energy consumption of the complete cell is $\approx  2.8$~pJ for the full cycle of the four memory operations, at a 50~\% duty cycle---that is, the case where the write process puts the skyrmion midway in the racetrack. This is the average case ($t=\tau/2$) of all possible temporal data recordings in the range $[0,  \tau ]$. 

The top panel of Fig.~\ref{fig:full_energy} shows that the energy consumption in the racetracks themselves (due to Joule heating) is only a very minor component of the total energy consumption and is on the order of $\approx 45$~fJ. Most of the energy is consumed in the driving and synchronizing transistors, primarily due to the high overdrive voltage applied to them. One way to reduce energy consumption is to optimize the energy of the driving and synchronizing transistors. Increasing their size up decreases the resistance, which will allow large drive current flow through the racetracks. Though the energy consumption of the transistors will decrease, energy consumption by the racetracks will increase. Even if increasing their size only moderately decreases the overall energy consumption, doing so would significantly improve the energy delay product. Further energy efficiency can be achieved if the necessary drive current for the skyrmion in a racetrack can be decreased without reducing the skyrmion velocity, as is the case for all racetrack memories. Specific optimization of these memory cells depends on the application and is left as a future exercise.

This memory cell is only useful if it has advantages over alternatives based solely on CMOS. One such alternative would be a up-counter coupled with a latch or an SRAM cell. The up-counter counts `clock ticks' and can thereby digitize a clock delay. An up-counter is built out of multiple positive edge-triggered D-flip flops (DFF) and combinational circuits. Textbook implementations of one such DFF requires 20 MOSFETs. Our racetracks store `analog' temporal information, while a counter stores quantized information. For example, a 6 stage counter, which has a $2^6 = 64$ step quantization of temporal data,  will require 120 transistors. Coupling them with simple S-R latches to store the memory requires another 24 transistors, yielding a total of 144 transistors. 

In addition, this scheme stores the data logarithmically and its readout requires either a Boolean decoder circuit increasing the number of transistors (an estimate for a $6\times64$ decoder requires 128 transistors) or a clock generator driven by latch readout (component count dependent on implementation scheme). Factoring in the built-in non-volatility of the skyrmion, where the proposed cell can be powered off without losing information for years, compared to the volatile nature of a pure CMOS design, makes it clear that the proposed memory cell is ultra-compact and would consume significantly less energy than such a CMOS-only design.

%\begin{figure}[t!]
%\centering
%\includegraphics[width=\linewidth]{figs/Picture3.png}
%\caption{Micromagnetic energy vs delay for 4 pairs of racetrack compared with a 4-bit CMOS counter. For lower damping proposed magnetic racetrack can be more efficient than its CMOS counterpart.}
%\label{energydelay}
%\end{figure}
%\begin{figure}[t!]
%\centering
%\includegraphics[width=\linewidth]{figs/pic3.png}
%\caption{Energy vs current density and comparing micromagnetic and analytical which show very good agreement.}
%\label{energy}
%\end{figure}

\subsection{Non-Idealities of Racetracks and their Possible Amelioration}

Our simulations are based on pristine racetracks, that is, there are no defects, particularly pinning centers. In experimental realizations, racetracks will suffer from a number of complications. In this section, we discuss these complications and possible techniques to minimize their impact.

\textbf{Defects: }Defects such as notches or material non-uniformity during fabrication of the racetrack give rise to pinning centers for skyrmions. The pinning and unpinning of skyrmions adds nonlinear effects to skyrmion movement described by the Eq.~\eqref{eq:simmobility}, giving rise to phenomena like creep.  Domain walls, another candidate for this approach, also suffer from pinning, but the two types of magnetic textures are more strongly affected by different types of disorder.  Since skyrmions are localized to the center of the racetrack, they are less susceptible to pinning by edge roughness than domain walls. On the other hand, because domain walls are more extended, they are less susceptible to pinning by the anisotropic grains that can exist in a racetrack~\cite{tomasello_performance_2017}. 

For low drive currents, the motion of skyrmions and domain walls are in the creep regime, where the motion consists of repeated pinning and thermally assisted depinning~\cite{lin2013particle}. The resulting stochastic motion of the magnetic textures causes errors in the spatial mapping of the temporal data as the skyrmion velocity is no longer a fully controllable parameter. While it is obvious that better fabrication of racetracks can ameliorate this unpredictability to an extent, we expect the presence of a residual density of defects under the best of the processes. Higher drive currents provide more energy to the motion, limiting defect based pinning and providing more predictable motion. However, higher currents increase the energy cost.

\textbf{Higher order skyrmion dynamics:}
At higher speeds the assumption of skyrmion rigidity is no longer valid. This can lead to nonlinear effects in skyrmion velocity~\cite{rigid}. For larger skyrmions, the existence of an inertia has been reported~\cite{inertia} which can effect the position of skyrmions. The breathing modes of skyrmions~\cite{breath} can also effect the motion, as it can change size and domain wall angle of a skyrmion, which in turn can introduce nonlinear effects to the skyrmion motion.

%\textbf{High speed limit of skyrmions:} It has been found \textcolor{cyan}{(\cite{kim_propulsion_2014,jin_dynamics_2016,salimath_controlling_2020})} that skyrmions show an in-built special relativity like effect with a maximal possible speed that is achieved asymptotically. This means that Eq.~\eqref{eq:simmobility} is valid only at low to moderate speeds \textcolor{cyan}{(under $\sim1km/s$)} before the relativistic effects are achieved \textcolor{red}{I suspect that equation will break down well before the relativistic limit is reached; I think the rigid skyrmion approximation is likely far more problematic than the fact that we're working at submagnonic speeds. In fact, I would just add a new subsection here called "higher order skyrmion dynamics" that discusses breathing modes, etc, and either remove this relativistic part or make it separate. If we do keep around a relativistic part, might be worth mentioning that the same physics happens in AFM DWs~\cite{kim_propulsion_2014} -MWD}. Since this is a fundamental property of skyrmion kinetics, it has to be taken into consideration for maximal drive current and bias points. The velocities we have worked with in this simulations are well below the relativistic limits.

\textbf{Skyrmion readability:} The read-out of skyrmion motion is through the TMR effect of the MTJ as mentioned before. Therefore the relative area of the skyrmion under the MTJ is a critical factor for obtaining a sufficiently large resistance swing to control the output swing of signal $V_M$ in Fig. 3. This matching of the area can be achieved by using large skyrmions to fill a significant area under the MTJ reader. Using domain walls removes this problem as they can be large enough to provide nearly full MTJ cross-section coverage. Improving the TMR of MTJ's would allow smaller skyrmions to be read with sufficient contrast. Another solution is to use additional transistors at the read stage to amplify the smaller voltage contrast provided by a low read TMR, but with a higher area and energy cost.

\textbf{Skyrmion lifetime:} Skyrmion lifetimes dictate the storage times of the memory cell. They depend on the skyrmion diameter and larger area skyrmions are more stable against annihilation \cite{bessarab_lifetime_2018}. Here, we consider skyrmions of diameter $48$~nm  with lifetimes of years as mentioned above. While in general smaller skyrmions might be preferable for high density binary memory, for a specialized application like this one, the skrmion size is less important, as the memory density is governed by the racetrack size, rather than the skyrmion size. The use of larger skyrmions also helps achieve better fill factors for readability and avoids accidental off-center injected skyrmions bypassing the MTJ altogether.

\textbf{Edge annihilation:} Skyrmions are susceptible to annihilation by the edges of the racetrack. This can be partially ameliorated by use of wide racetracks, whereby we can avoid the possibility of skyrmion drifting to the edges and getting annihilated due to Magnus effect. This however can worsen the skyrmion readability as this scales up the size of the read MTJs, or makes skyrmions miss the MTJ as discussed above. While the use of synthetic antiferromagnet racetracks with compensation for Magnus force can considerably reduce this issue, use of reflecting edges through anisotropy engineering, such as using ion beam irradiation, or geometry engineering by using thicker edges can also help with this issue\cite{fook_mitigation_2015,fallon_controlled_2020}. By using reflective edges, the skyrmion will more reliably reach the MTJ position, as the repulsive force from the edges will confine the skyrmion to the middle of racetrack.

All of these issues listed above are common to the area of magnetic racetracks, an area of active research, and any future improvements through better fabrication capabilities and novel extrinsic circuit techniques will benefit the performance and shortcomings of the proposed temporal memory cell.

\section{Conclusion}
\label{sec:conclusion}

Faced with the unrelenting demand of tomorrow's computing needs, coupled with ending of Moore's law, engineers are faced with the task of questioning some of the fundamental assumptions that have driven the development of conventional computing systems for many years. While changing the information representation may be a small step in the quest for efficiency, it requires the development of supporting systems like memory that can naturally encode information in the same domain, so as to cut domain translation costs. 

In this work, we utilize a novel technology based on skyrmions in magnetic racetracks to provide energy-efficient memory for temporal computing. Our proposal uses the displacement of pre-nucleated skyrmions to store temporal information in magnetic racetracks, avoiding the energy cost required to nucleate them. The readout of this memory is made non-destructive by doubling the number of racetracks to store the information in the second racetrack during the read process. While this increases the energy consumption and area of the memory, we limit the impact by sharing readout and control circuits for the different operations involved in using the memory. The linearity of the stored information with respect to the input limits the circuitry required for translation between the time and displacement domains.  By changing the current used to drive the skyrmions, and hence their velocity, these memory cells can be tuned so their time scale matches the range of input information. Such an efficient memory will greatly expand the problem domain that can be efficiently addressed by timing based computing.  

\section{Acknowledgments}
\label{sec:acknowlege}
This work is funded in part by the Defense Advanced Research Projects Agency (DARPA) Topological Excitations in Electronics (TEE) program (grant  D18AP00009). A.M. acknowledges support under the Cooperative Research Agreement Award No.~70NANB14H209, through the University  of Maryland. We would like to thank Andrew Kent, Joe Poon, Geoffrey Beach, and Brian Hoskins for insightful discussions.

\balance

%\bibliographystyle{IEEEtran}
%\bibliography{IEEEabrv,mybibfile}

%% Generated by IEEEtran.bst, version: 1.14 (2015/08/26)

% biography section
% 
% If you have an EPS/PDF photo (graphicx package needed) extra braces are
% needed around the contents of the optional argument to biography to prevent
% the LaTeX parser from getting confused when it sees the complicated
% \includegraphics command within an optional argument. (You could create
% your own custom macro containing the \includegraphics command to make things
% simpler here.)
%\begin{IEEEbiography}[{\includegraphics[width=1in,height=1.25in,clip,keepaspectratio]{mshell}}]{Michael Shell}
% or if you just want to reserve a space for a photo:

%\begin{IEEEbiography}{Michael Shell}
%Biography text here.
%\end{IEEEbiography}

% if you will not have a photo at all:
%\begin{IEEEbiographynophoto}{John Doe}
%Biography text here.
%\end{IEEEbiographynophoto}

% insert where needed to balance the two columns on the last page with
% biographies
%\newpage

%\begin{IEEEbiographynophoto}{Jane Doe}
%Biography text here.
%\end{IEEEbiographynophoto}

% You can push biographies down or up by placing
% a \vfill before or after them. The appropriate
% use of \vfill depends on what kind of text is
% on the last page and whether or not the columns
% are being equalized.

%\vfill

% Can be used to pull up biographies so that the bottom of the last one
% is flush with the other column.
%\enlargethispage{-5in}

% that's all folks
\end{document}